\documentclass[aps,secnumarabic,nobalancelastpage,
nofootinbib,onecolumn,12pt]{revtex4}%
\usepackage{amsfonts}
\usepackage{amsmath}
\usepackage{graphics}
\usepackage{graphicx}
\usepackage{longtable}
\usepackage{url}
\usepackage{bm}
\usepackage{amssymb}%
\begin{document}
\title{Analysis of Generalized Debye-H\"{u}ckel Equation from Poisson-Fermi Theory }
\author{Chin-Lung Li and Jinn-Liang Liu}
\affiliation{Institute of Computational and Modeling Science, National Tsing Hua
University, Hsinchu 300, Taiwan. E-mail: chinlungli@mail.nd.nthu.edu.tw, jlliu@mx.nthu.edu.tw}
\date{\today }

\begin{abstract}

\end{abstract}
\maketitle

\textbf{Abstract.} The Debye-H\"{u}ckel equation is a fundamental physical
model in chemical thermodynamics that describes the free energy (chemical
potential, activity) of an ion in electrolyte solutions at variable salt
concentration, temperature, and pressure. It is based on the linear
Poisson-Boltzmann equation that ignores the steric (finite size), correlation,
and polarization effects of ions and water (or solvent molecules). The
Poisson-Fermi theory developed in recent years takes these effects into
account. A generalized Debye-H\"{u}ckel equation is derived from the
Poisson-Fermi theory and is shown to consistently reduce to the classical
equation when these effects vanish in limiting cases. As a result, a linear
fourth-order Poisson-Fermi equation is presented for which unique solutions
are shown to exist for spherically symmetric systems. Moreover, a generalized
Debye length is proposed to include the size effects of ions and water.

\section{Introduction}

Thermodynamic modeling of aqueous electrolyte solutions plays a fundamental
role in chemical and biological sciences and engineering
\cite{BK09,E13,F04,H01,KF09,KM18,K10,LM03,LJ08,N91,P95,RK15,RS59,VW16,V11}.
Despite intense efforts in the past century, robust thermodynamic modeling of
electrolyte solutions still remains a remote ambition \cite{RK15} in the
extended models from the classical Debye-H\"{u}ckel (DH) theory due to the
enormous number of parameters that need to be adjusted, carefully and often
subjectively \cite{F10,KM18,RK15,V11}. For example, the Pitzer model requires
8 parameters for a ternary system and up to 8 temperature coefficients
(parameters) for every Pitzer parameter in a temperature interval from 0 to
about 200 $%
\operatorname{{}^{\circ}{\rm C}}%
$ \cite{RK15,V11}. It is indeed a frustrating despair (\textit{frustration} on
p. 11 in \cite{K10} and \textit{despair }on p. 301 in \cite{RS59}) that
approximately 22,000 parameters for combinatorial solutions of the most
important 28 cations and 16 anions in salt chemistry have to be extracted from
the available experimental data for one temperature \cite{V11}. The Pitzer
model is still the most widely used DH model with unmatched precision for
modeling aqueous electrolyte solutions over wide ranges of composition,
temperature, and pressure \cite{RK15}.

The Debye-H\"{u}ckel theory \cite{DH23} is based on a linearization of the
nonlinear Poisson-Boltzmann (PB) equation that was developed by Gouy
\cite{G10} and Chapman \cite{C13} in early 1910s and ignores the steric
(finite size), correlation, and polarization effects of ions and water.
However, tremendous technological advances of experimental tools in modern
sciences render physical details of biological and chemical systems at atomic
scale \cite{B00,GM09} for which these effects can no longer be ignored for
modeling electrolyte solutions in numerous applications
\cite{BK09,E13,F04,KM18}. The Poisson-Fermi (PF) theory developed in recent
years takes these effects into account and has been shown to illustrate the
importance of these effects in a range of areas from electric double layers
\cite{BS11,L13,LX17} and ion activities \cite{LE15a,LE18} to biological ion
channels \cite{L13,LE14,LE15,LH16,LX17}.

Based on the numerical PF model developed in \cite{LE15a,LE18}, we derive,
analyze, and verify a generalized DH equation from the PF model to account for
these three effects. The PF model is a fourth-order nonlinear partial
differential equation (PDE), where the fourth-order term describes the
correlation effect of ions and the polarization effect of water molecules both
in a mean-field approximation. The finite-size (steric) effect of all
particles (ions and water treated as non-uniform spheres) is described by a
steric potential that is defined by a void fraction function in the solvent
domain, i.e., the voids between spheres are expressed as a function that
varies in the domain with the variation of electric potential if the solvent
domain is subject to external electric fields. The steric potential can be
considered as a mean-field approximation of the Lennard-Jones (L-J) potential
between any pair of ions and water molecules, which describes the net
inter-molecular force and thus defines the distance (voidness) between the
pair. L-J potentials are highly oscillatory and hence not suitable for
mean-field calculations due to approximation and convergence issues
\cite{L13,LE15}.

The main results of the present work are summarized as follows: (i) A
fourth-order lineal PF equation is presented for which we show that unique
solutions exist for spherical symmetric systems such as the ionic activity
model considered here. (ii) A generalized Debye length --- a measure of how
far the electrostatic effect of an ion in solution persists \cite{DH23} --- is
proposed to include the size effects of ions and water molecules, which have
been ignored in the classical Debye length. (iii) A correlation length
\cite{LF96} --- a measure of how strongly correlations between ions develop
\cite{BS11,S06} \textit{and} how easily water molecules in electrolyte
solutions are polarized in response to an electric field
\cite{L13,LE13,LE14,LE15} --- is shown to be a \textit{non}-empirical
parameter (in contrast to empirical ones in \cite{BS11,L13,LE14,LE15,S06})
that depends not only on the salt concentration but also on the sizes of ions
and water (in contrast to that of size independence in \cite{LF96}). (iv) The
generalized DH equation and Debye length are shown to reduce to their
classical versions when the steric and correlation effects are ignored.

The remaining of the paper is organized as follows. The Poisson-Fermi theory
is briefly described in Section 2 from which we give a detailed derivation and
analysis of the generalized Debye-H\"{u}ckel equation in Section 3. The
derivation is based on a linear Poisson-Fermi equation in Section 3.1, which
is a fourth-order PDE and is transformed to two second-order PDEs in Section
3.2 for which general solutions are found for spherical symmetric systems. In
Section 3.3, a unique solution is found for the second-order PDEs in a
specific domain that consists of effective Born, hydration shell, and solvent
spherical subdomains for modeling the solvation energy of an ion in general
binary electrolytes. We present a full set of interface and boundary
conditions for these two PDEs on which the unique potential solution is
derived. This electrostatic potential accounts for correlation, steric,
polarization, and hydration effects that are absent in the classical DH
theory. It also leads to a generalized DH equation in Section 3.4 for
algebraically calculating individual or mean activity coefficients of binary
electrolytes. In Section 3.4, we show that the generalized equation
consistently reduces to the classical DH equation when these effects are
ignored. Moreover, asymptotic analyses of the linear PF equation (Section 3.2)
and the generalized DH equation (Section 3.4) are also given as the
generalized Debye length tends to infinite (or equivalently the salt
concentration tends to zero) for infinite dilute electrolytes, the correlation
length tends to zero (without correlations), or the volumes of ions and water
tend to zero (without size effects). Some concluding remarks of this work are
made in Section 4.

\section{Poisson-Fermi Theory}

For an aqueous electrolyte system with $K$ species of ions, the entropy model
proposed in \cite{LE14,LX17} treats all ions and water of any diameter as
nonuniform hard spheres, and regards the water as the $(K+1)^{\text{th}}$
species and the voids between these hard spheres as the $(K+2)^{\text{th}}$
species. The total volume $V$ of the system can be calculated exactly by the
identity%
\begin{equation}
V=\sum_{i=1}^{K+1}v_{i}N_{i}+V_{K+2}, \label{2.1}%
\end{equation}
where $v_{i}=4\pi a_{i}^{3}/3$ with radius $a_{i}$, $N_{i}$ is the total
number of the $i^{\text{th}}$ species particles, and $V_{K+2}$ denotes the
total volume of all the voids. In the bulk solution, we have the bulk
concentrations $C_{i}^{B}=\frac{N_{i}}{V}$ and the bulk volume fraction of
voids $\Gamma^{B}=\frac{V_{K+2}}{V}$. Dividing the volume identity (\ref{2.1})
by $V$, $\Gamma^{B}=1-\sum_{i=1}^{K+1}v_{i}C_{i}^{B}$ is expressed in terms of
nonuniform $v_{i}$ and $C_{i}^{B}$ for all particle species. If the system is
spatially inhomogeneous with variable electric or steric fields, as in
realistic biological and chemical systems, the bulk concentrations then change
to concentration functions $C_{i}(\mathbf{r})$ that vary with positions, and
differ from their constant values $C_{i}^{B}$ at location $\mathbf{r}$ in the
solvent domain $\Omega_{s}$. Consequently, the void volume fraction becomes a
function $\Gamma(\mathbf{r)}=1-\sum_{i=1}^{K+1}v_{i}C_{i}(\mathbf{r})$ as well.

It is shown in \cite{LE14,LX17} that the distribution (concentration) of
particles in $\Omega_{s}$ is of Fermi-like type%
\begin{equation}
C_{i}(\mathbf{r})=C_{i}^{B}\exp\left(  -\beta_{i}\phi(\mathbf{r})+\frac{v_{i}%
}{v_{0}}S^{\text{trc}}(\mathbf{r})\right)  \text{, \ \ }S^{\text{trc}%
}(\mathbf{r})=\ln\left(  \frac{\Gamma(\mathbf{r)}}{\Gamma^{B}}\right)  ,
\label{2.2}%
\end{equation}
since it saturates, i.e., $C_{i}(\mathbf{r})<\frac{1}{v_{i}}$ for any
arbitrary (or even infinite) electric potential $\phi(\mathbf{r})$ at any
location $\mathbf{r\in}$ $\Omega_{s}$ for all $i=1,$ $\cdots,$ $K+1$ (ions and
water), where $\beta_{i}=q_{i}/k_{B}T$ with $q_{i}$ being the charge on
species $i$ particles and $q_{K+1}=0$, $k_{B}$ is the Boltzmann constant, $T$
is an absolute temperature, and $v_{0}=\left(  \sum_{i=1}^{K+1}v_{i}\right)
/(K+1)$ an average volume. The steric potential $S^{\text{trc}}(\mathbf{r})$
first proposed in \cite{L13} is an entropic measure of crowding or emptiness
of particles at $\mathbf{r}$. If $\phi(\mathbf{r})=0$ for all $\mathbf{r}$,
then $\Gamma(\mathbf{r)}=\Gamma^{\text{B}}$ and hence $S^{\text{trc}%
}(\mathbf{r})=0$. The factor $v_{i}/v_{0}$ in (\ref{2.2}) shows that the
steric energy $\frac{-v_{i}}{v_{0}}S^{\text{trc}}(\mathbf{r})k_{B}T$ of a type
$i$ particle at $\mathbf{r}$ depends not only on the steric potential
$S^{\text{trc}}(\mathbf{r})$ but also on its volume $v_{i}$ similar to the
electric energy $\beta_{i}\phi(\mathbf{r})k_{B}T$ that depends on both the
electric potential $\phi(\mathbf{r})$ and its charge $q_{i}$ \cite{LX17}. The
steric potential is a mean-field approximation of Lennard-Jones (L-J)
potentials that describe local variations of L-J distances (and thus empty
voids) between every pair of particles. L-J potentials are highly oscillatory
and extremely expensive and unstable to compute numerically.

A nonlocal electrostatic formulation of ions and water is proposed in
\cite{LX17} to describe the correlation effect of ions and the polarization
effect of polar water. The formulation yields the following fourth-order
Poisson-Fermi equation \cite{S06}%
\begin{equation}
\epsilon_{s}\left(  l_{c}^{2}\nabla^{2}-1\right)  \nabla^{2}\phi
(\mathbf{r})=\rho(\mathbf{r}),\text{\ }\mathbf{r}\in\Omega_{s}\text{,}
\label{2.3}%
\end{equation}
where $\epsilon_{s}=\epsilon_{w}\epsilon_{0}$, $\epsilon_{w}$ is the
dielectric constant of bulk water, $\epsilon_{0}$ is the vacuum permittivity,
$l_{c}=\sqrt{l_{B}l_{D}/48}$ is a density-density correlation length
independent of specific ionic radius \cite{LF96}, $l_{B}$ and $l_{D}$ are the
Bjerrum and Debye lengths, respectively, $\nabla$ is the gradient operator in
$R^{3}$, $\nabla^{2}=\nabla\cdot\nabla$, and $\rho(\mathbf{r})=\sum_{i=1}%
^{K}q_{i}C_{i}(\mathbf{r})$ is ionic charge density. It is shown in
\cite{LX17} that the fourth-order PF equation (\ref{2.3}) reduces to the
classical second-order PB equation when $l_{c}=v_{i}=0$ for all $i$, i.e., the
correlation, steric, and polarization effects are ignored. Eq. (\ref{2.3}) was
first proposed in \cite{S06} (with $v_{i}=0$ for all $i$) and subsequently
derived in \cite{BS11,LX17} (with $v_{i}\neq0$) from different perspectives of electrostatics.

\section{Derivation and Analysis of Generalized Debye-H\"{u}ckel Equation}

\subsection{Linear Poisson-Fermi Equation}

For simplicity, we consider a general binary ($K=2$) electrolyte C$_{z_{2}}%
$A$_{z_{1}}$ with the valences of the cation C$^{z_{1}+}$ and anion
A$^{z_{2}-} $ being $z_{1}$ and $z_{2}$, respectively. In the bulk situation
($\phi(\mathbf{r})=S^{\text{trc}}(\mathbf{r})=0$), the charge neutrality
condition $q_{1}N_{1}+q_{2}N_{2}=0$ of the system implies that%
\begin{align*}
\beta_{2}  &  =\frac{q_{2}}{k_{B}T}=\frac{-N_{1}q_{1}}{N_{2}k_{B}T}%
=\frac{-N_{1}}{N_{2}}\beta_{1}=\frac{-C_{1}^{B}}{C_{2}^{B}}\beta_{1},\\
C_{2}^{B}  &  =\frac{N_{2}}{V}=\frac{-q_{1}N_{1}}{q_{2}V}=\frac{-q_{1}%
C_{1}^{B}}{q_{2}},
\end{align*}
and hence%
\begin{equation}
\epsilon_{s}\left(  l_{c}^{2}\nabla^{2}-1\right)  \nabla^{2}\phi
(\mathbf{r})=q_{1}C_{1}(\mathbf{r})+q_{2}C_{2}(\mathbf{r})=q_{1}\left[
C_{1}(\mathbf{r})-\frac{C_{1}^{B}}{C_{2}^{B}}C_{2}(\mathbf{r})\right]  .
\label{3.1}%
\end{equation}
Since%
\begin{align*}
C_{i}(\mathbf{r})  &  =C_{i}^{B}\exp\left(  -\beta_{i}\phi(\mathbf{r})\right)
\left(  \frac{\Gamma(\mathbf{r)}}{\Gamma^{B}}\right)  ^{\frac{v_{i}}{v_{0}}%
}\text{, }i=1,\text{ }2\text{,}\\
C_{3}(\mathbf{r})  &  =C_{3}^{B}\left(  \frac{\Gamma(\mathbf{r)}}{\Gamma^{B}%
}\right)  ^{\frac{v_{3}}{v_{0}}}\text{,}%
\end{align*}
we obtain%
\begin{align}
\frac{\Gamma(\mathbf{r)}}{\Gamma^{B}}  &  =\left(  \frac{C_{3}(\mathbf{r}%
)}{C_{3}^{B}}\right)  ^{\frac{v_{0}}{v_{3}}},\label{3.2}\\
C_{i}(\mathbf{r})  &  =C_{i}^{B}\exp\left(  -\beta_{i}\phi(\mathbf{r})\right)
\left(  \frac{C_{3}(\mathbf{r})}{C_{3}^{B}}\right)  ^{\frac{v_{i}}{v_{3}}%
}\text{.} \label{3.3}%
\end{align}
Substituting Eq. (\ref{3.3}) into (\ref{3.2}) yields%
\begin{align}
\left(  \frac{C_{3}(\mathbf{r})}{C_{3}^{B}}\right)  ^{\frac{v_{0}}{v_{3}}}  &
=\frac{1-v_{1}C_{1}^{B}\exp\left(  -\beta_{1}\phi(\mathbf{r})\right)  \left(
\frac{C_{3}(\mathbf{r})}{C_{3}^{B}}\right)  ^{\frac{v_{1}}{v_{3}}}}{\Gamma
^{B}}\nonumber\\
&  -\frac{v_{2}C_{2}^{B}\exp\left(  -\beta_{2}\phi(\mathbf{r})\right)  \left(
\frac{C_{3}(\mathbf{r})}{C_{3}^{B}}\right)  ^{\frac{v_{2}}{v_{3}}}+v_{3}%
C_{3}(\mathbf{r})}{\Gamma^{B}}. \label{3.4}%
\end{align}
Assuming that the functional $C_{3}(\phi(\mathbf{r}))=C_{3}(\mathbf{r})$ can
be expressed by Taylor's formula%
\begin{equation}
C_{3}(\mathbf{r})=b_{0}+b_{1}\phi(\mathbf{r})+O(\phi^{2}(\mathbf{r})),
\label{3.5}%
\end{equation}
we then have%
\begin{align}
\left(  \frac{C_{3}(\mathbf{r})}{C_{3}^{B}}\right)  ^{\alpha}  &  =\left(
\frac{b_{0}}{C_{3}^{B}}\right)  ^{\alpha}+\alpha\left(  \frac{b_{1}}{C_{3}%
^{B}}\right)  \left(  \frac{b_{0}}{C_{3}^{B}}\right)  ^{\alpha-1}%
\phi(\mathbf{r})+O(\phi^{2}(\mathbf{r}))\nonumber\\
&  =\left(  \frac{b_{0}}{C_{3}^{B}}\right)  ^{\alpha}+\alpha\left(
\frac{b_{1}}{b_{0}}\right)  \left(  \frac{b_{0}}{C_{3}^{B}}\right)  ^{\alpha
}\phi(\mathbf{r})+O(\phi^{2}(\mathbf{r}))\text{ for }\alpha\geq0. \label{3.6}%
\end{align}
Consequently, the left hand side of Eq. (\ref{3.4}) can be written as%
\begin{equation}
\left(  \frac{C_{3}(\mathbf{r})}{C_{3}^{B}}\right)  ^{\frac{v_{0}}{v_{3}}%
}=\left(  \frac{b_{0}}{C_{3}^{B}}\right)  ^{\frac{v_{0}}{v_{3}}}+\frac{v_{0}%
}{v_{3}}\frac{b_{1}}{b_{0}}\left(  \frac{b_{0}}{C_{3}^{B}}\right)
^{\frac{v_{0}}{v_{3}}}\phi(\mathbf{r})+O(\phi^{2}(\mathbf{r})) \label{3.7}%
\end{equation}
and the right hand side gives%
\begin{align}
&  \frac{1-v_{1}C_{1}^{B}\exp\left(  -\beta_{1}\phi(\mathbf{r})\right)
\left(  \frac{C_{3}(\mathbf{r})}{C_{3}^{B}}\right)  ^{\frac{v_{1}}{v_{3}}%
}-v_{2}C_{2}^{B}\exp\left(  -\beta_{2}\phi(\mathbf{r})\right)  \left(
\frac{C_{3}(\mathbf{r})}{C_{3}^{B}}\right)  ^{\frac{v_{2}}{v_{3}}}-v_{3}%
C_{3}(\mathbf{r})}{\Gamma^{B}}\nonumber\\
&  =\frac{1-v_{1}C_{1}^{B}\left(  1-\beta_{1}\phi(\mathbf{r})+O(\phi
^{2}(\mathbf{r}))\right)  \left[  \left(  \frac{b_{0}}{C_{3}^{B}}\right)
^{\frac{v_{1}}{v_{3}}}+\frac{v_{0}}{v_{3}}\frac{b_{1}}{b_{0}}\left(
\frac{b_{0}}{C_{3}^{B}}\right)  ^{\frac{v_{1}}{v_{3}}}\phi(\mathbf{r}%
)+O(\phi^{2}(\mathbf{r}))\right]  }{\Gamma^{B}}\nonumber\\
&  -\frac{v_{2}C_{2}^{B}\left(  1-\beta_{2}\phi(\mathbf{r})+O(\phi
^{2}(\mathbf{r}))\right)  \left[  \left(  \frac{b_{0}}{C_{3}^{B}}\right)
^{\frac{v_{2}}{v_{3}}}+\frac{v_{0}}{v_{3}}\frac{b_{1}}{b_{0}}\left(
\frac{b_{0}}{C_{3}^{B}}\right)  ^{\frac{v_{2}}{v_{3}}}\phi(\mathbf{r}%
)+O(\phi^{2}(\mathbf{r}))\right]  }{\Gamma^{B}}\nonumber\\
&  -\frac{v_{3}\left(  b_{0}+b_{1}\phi(\mathbf{r})+O(\phi^{2}(\mathbf{r}%
))\right)  }{\Gamma^{B}}\nonumber\\
&  =\frac{1-v_{1}C_{1}^{B}\left(  \frac{b_{0}}{C_{3}^{B}}\right)
^{\frac{v_{1}}{v_{3}}}-v_{2}C_{2}^{B}\left(  \frac{b_{0}}{C_{3}^{B}}\right)
^{\frac{v_{2}}{v_{3}}}-v_{3}b_{0}}{\Gamma^{B}}\nonumber\\
&  +\frac{v_{1}C_{1}^{B}\left(  \frac{b_{0}}{C_{3}^{B}}\right)  ^{\frac{v_{1}%
}{v_{3}}}\left[  \beta_{1}-\left(  \frac{v_{0}}{v_{3}}\right)  \left(
\frac{b_{1}}{b_{0}}\right)  \right]  }{\Gamma^{B}}\phi(\mathbf{r})\nonumber\\
&  +\frac{v_{2}C_{2}^{B}\left(  \frac{b_{0}}{C_{3}^{B}}\right)  ^{\frac{v_{2}%
}{v_{3}}}\left[  \beta_{2}-\left(  \frac{v_{0}}{v_{3}}\right)  \left(
\frac{b_{1}}{b_{0}}\right)  \right]  -v_{3}b_{1}}{\Gamma^{B}}\phi
(\mathbf{r})+O(\phi^{2}(\mathbf{r})). \label{3.8}%
\end{align}
Therefore, the constant terms in Eqs. (\ref{3.7}) and (\ref{3.8}) give%
\begin{equation}
\left(  \frac{b_{0}}{C_{3}^{B}}\right)  ^{\frac{v_{0}}{v_{3}}}=\frac
{1-v_{1}C_{1}^{B}\left(  \frac{b_{0}}{C_{3}^{B}}\right)  ^{\frac{v_{1}}{v_{3}%
}}-v_{2}C_{2}^{B}\left(  \frac{b_{0}}{C_{3}^{B}}\right)  ^{\frac{v_{2}}{v_{3}%
}}-v_{3}b_{0}}{\Gamma^{B}} \label{3.9}%
\end{equation}
whereas the first-order terms yield%
\begin{align}
\frac{v_{0}b_{1}}{v_{3}b_{0}}\left(  \frac{b_{0}}{C_{3}^{B}}\right)
^{\frac{v_{0}}{v_{3}}}  &  =\frac{v_{1}C_{1}^{B}\left(  \frac{b_{0}}{C_{3}%
^{B}}\right)  ^{\frac{v_{1}}{v_{3}}}\left(  \beta_{1}-\frac{v_{1}b_{1}}%
{v_{3}b_{0}}\right)  }{\Gamma^{B}}\nonumber\\
&  +\frac{v_{2}C_{2}^{B}\left(  \frac{b_{0}}{C_{3}^{B}}\right)  ^{\frac{v_{2}%
}{v_{3}}}\left(  \beta_{2}-\frac{v_{2}b_{1}}{v_{3}b_{0}}\right)  -v_{3}b_{1}%
}{\Gamma^{B}}. \label{3.10}%
\end{align}
Note that $b_{0}=C_{3}^{B}$ is a solution of Eq. (\ref{3.9}). To determine if
this solution is unique, we define the function%
\[
f(x)=\Gamma^{B}x^{\frac{v_{0}}{v_{3}}}+v_{1}C_{1}^{B}x^{\frac{v_{1}}{v_{3}}%
}+v_{2}C_{2}^{B}x^{\frac{v_{2}}{v_{3}}}+v_{3}C_{3}^{B}x-1
\]
that gives%
\[
\lim_{x\rightarrow0^{+}}f(x)=-1\text{, }\lim_{x\rightarrow\infty}%
f(x)=\infty>0\text{,}%
\]%
\[
f^{\prime}(x)=\frac{v_{0}}{v_{3}}\Gamma^{B}x^{\frac{v_{0}}{v_{3}}-1}%
+\frac{v_{1}}{v_{3}}v_{1}C_{1}^{B}x^{\frac{v_{1}}{v_{3}}-1}+\frac{v_{2}}%
{v_{3}}v_{2}C_{2}^{B}x^{\frac{v_{2}}{v_{3}}-1}+v_{3}C_{3}^{B}>0.
\]
Therefore, the coefficient $b_{0}=C_{3}^{B}$ is unique. Similarly, by Eq.
(\ref{3.10}), we have%
\begin{align*}
\frac{v_{0}b_{1}}{v_{3}C_{3}^{B}}  &  =\frac{v_{1}C_{1}^{B}\left(  \beta
_{1}-\frac{v_{1}b_{1}}{v_{3}C_{3}^{B}}\right)  +v_{2}C_{2}^{B}\left(
\beta_{2}-\frac{v_{2}b_{1}}{v_{3}C_{3}^{B}}\right)  -v_{3}b_{1}}{\Gamma^{B}}\\
&  =\frac{\beta_{1}C_{1}^{B}\left(  v_{1}-v_{2}\right)  -\frac{\left(
v_{1}^{2}C_{1}^{B}+v_{2}^{2}C_{2}^{B}\right)  b_{1}}{v_{3}C_{3}^{B}}%
-v_{3}b_{1}}{\Gamma^{B}}\text{,}%
\end{align*}%
\[
\Gamma^{B}v_{0}b_{1}=\beta_{1}C_{1}^{B}C_{3}^{B}v_{3}\left(  v_{1}%
-v_{2}\right)  -\left(  v_{1}^{2}C_{1}^{B}+v_{2}^{2}C_{2}^{B}+v_{3}^{2}%
C_{3}^{B}\right)  b_{1}\text{,}%
\]
and hence the coefficient $b_{1}=\frac{\beta_{1}C_{1}^{B}C_{3}^{B}v_{3}\left(
v_{1}-v_{2}\right)  }{\Gamma^{B}v_{0}+\left(  v_{1}^{2}C_{1}^{B}+v_{2}%
^{2}C_{2}^{B}+v_{3}^{2}C_{3}^{B}\right)  }$ is unique by Eqs. (\ref{3.9}) and
(\ref{3.10}). From Eqs. (\ref{3.1}), (\ref{3.3}), and (\ref{3.6}), we have
\begin{align*}
&  q_{1}\left[  C_{1}(\mathbf{r})-\frac{C_{1}^{B}}{C_{2}^{B}}C_{2}%
(\mathbf{r})\right] \\
&  =q_{1}C_{1}^{B}\left[  \exp\left(  -\beta_{1}\phi(\mathbf{r})\right)
\left(  \frac{C_{3}(\mathbf{r})}{C_{3}^{B}}\right)  ^{\frac{v_{1}}{v_{3}}%
}-\exp\left(  -\beta_{2}\phi(\mathbf{r})\right)  \left(  \frac{C_{3}%
(\mathbf{r})}{C_{3}^{B}}\right)  ^{\frac{v_{2}}{v_{3}}}\right] \\
&  =q_{1}C_{1}^{B}\left\{  \left(  1-\beta_{1}\phi(\mathbf{r})+O(\phi
^{2}(\mathbf{r}))\right)  \left[  \left(  \frac{b_{0}}{C_{3}^{B}}\right)
^{\frac{v_{1}}{v_{3}}}+\frac{v_{1}b_{1}}{v_{3}b_{0}}\left(  \frac{b_{0}}%
{C_{3}^{B}}\right)  ^{\frac{v_{1}}{v_{3}}}\phi(\mathbf{r})+O(\phi
^{2}(\mathbf{r}))\text{ }\right]  \right. \\
&  -\left.  \left(  1-\beta_{2}\phi(\mathbf{r})+O(\phi^{2}(\mathbf{r}%
))\right)  \left[  \left(  \frac{b_{0}}{C_{3}^{B}}\right)  ^{\frac{v_{2}%
}{v_{3}}}+\frac{v_{2}b_{1}}{v_{3}b_{0}}\left(  \frac{b_{0}}{C_{3}^{B}}\right)
^{\frac{v_{2}}{v_{3}}}\phi(\mathbf{r})+O(\phi^{2}(\mathbf{r}))\text{ }\right]
\right\} \\
&  =q_{1}C_{1}^{B}\left\{  \left(  1-\beta_{1}\phi(\mathbf{r})+O(\phi
^{2}(\mathbf{r}))\right)  \left[  1+\frac{v_{1}b_{1}}{v_{3}C_{3}^{B}}%
\phi(\mathbf{r})+O(\phi^{2}(\mathbf{r}))\text{ }\right]  \right. \\
&  -\left.  \left(  1-\beta_{2}\phi(\mathbf{r})+O(\phi^{2}(\mathbf{r}%
))\right)  \left[  1+\frac{v_{2}b_{1}}{v_{3}C_{3}^{B}}\phi(\mathbf{r}%
)+O(\phi^{2}(\mathbf{r}))\text{ }\right]  \right\} \\
&  =q_{1}C_{1}^{B}\left[  \left(  \beta_{2}-\beta_{1}\right)  +\frac{\left(
v_{1}-v_{2}\right)  b_{1}}{v_{3}C_{3}^{B}}\right]  \phi(\mathbf{r})+O(\phi
^{2}(\mathbf{r}))\text{,}%
\end{align*}
which implies the following result.

\textbf{Theorem 3.1.} If Taylor's formula (\ref{3.5}) holds for the functional
$C_{3}(\phi(\mathbf{r}))=C_{3}(\mathbf{r})$, we then have the linear
Poisson-Fermi equation%
\begin{equation}
\epsilon_{s}\left(  l_{c}^{2}\nabla^{2}-1\right)  \nabla^{2}\phi
(\mathbf{r})=\frac{-C_{1}^{B}q_{1}}{k_{B}T}\left[  \left(  q_{1}-q_{2}\right)
-\Lambda q_{1}\right]  \phi(\mathbf{r}) \label{3.11}%
\end{equation}
for any binary ($K=2$) electrolyte solutions, where
\begin{equation}
\Lambda=\frac{C_{1}^{B}\left(  v_{1}-v_{2}\right)  ^{2}}{\Gamma^{B}%
v_{0}+\left(  v_{1}^{2}C_{1}^{B}+v_{2}^{2}C_{2}^{B}+v_{3}^{2}C_{3}^{B}\right)
}\text{.} \label{3.12}%
\end{equation}
Consequently, we obtain a generalized Debye length
\begin{equation}
l_{DPF}=\left(  \frac{\epsilon_{s}k_{B}T}{C_{1}^{B}((1-\Lambda)q_{1}^{2}%
-q_{1}q_{2})}\right)  ^{1/2} \label{3.13}%
\end{equation}

\textbf{Remark 3.1.} The generalized Debye length $l_{DPF}$ appears to be
first proposed in the literature to our knowledge, where $\Lambda$ is a
dimensionless quantity corresponding to the size effects. It depends not only
on the charges but also on the sizes of all particles (ions and water). By
contrast, the classical Debye length $l_{D}$ \cite{LM03} depends only on
charges but not sizes, since all particles were treated as volumeless points
in the classical Debye-H\"{u}ckel formulation of the linear Poisson-Boltzmann
equation \cite{LM03}. Note that the generalized length reduces to the
classical length if $v_{1}=v_{2}\neq0$ (two ionic species having equal radius
and thus $\Lambda=0$) or $v_{1}=v_{2}=$ $v_{3}=0$ (all particles are points).
The linear PF equation (\ref{3.11}) is simplified to%
\begin{equation}
\left(  1-l_{c}^{2}\nabla^{2}\right)  \nabla^{2}\phi(\mathbf{r})=\kappa
^{2}\phi(\mathbf{r}), \label{3.14}%
\end{equation}
where $\kappa^{2}=\frac{C_{1}^{B}q_{1}}{\epsilon_{s}k_{B}T}\left[  \left(
q_{1}-q_{2}\right)  -\Lambda q_{1}\right]  $ and $\kappa^{-1}=l_{DPF}$. It
reduces to the linear PB equation if both correlation ($l_{c}=0$) and steric
($\Lambda=0$) effects are ignored. Note that $l_{DPF}$ (or $l_{D}$) is
proportional to $\frac{1}{\sqrt{C_{1}^{B}}}$.

\textbf{Remark 3.2.} Since $l_{DPF}$ includes the size effect, the correlation
length $l_{c}$ should be generalized to $l_{c}=\sqrt{l_{B}l_{DPF}/48}$ for
taking this effect into account as well. Consequently, the generalized
correlation length is not \textit{universal} \cite{LF96} but size dependent in
contrast to that in \cite{LF96} derived from $l_{D}$. Correlation lengths used
in previous works \cite{BS11,L13,LE15a,LX17,S06} are all \textit{empirical}
constants that depend specifically on the size and/or valence of a particular
ion of interest. The correlation length used here is \textit{not }an empirical parameter.

\subsection{General Solutions of Linear PF Equation in Spherical Symmetric
System}

Analytical solutions of the linear PB equation are in general not available
for arbitrary domains except for special cases such as spherical domains
\cite{K62}. The classical DH equation was derived from the linear PB equation
in a spherically symmetric system \cite{LM03}. We now find a general solution
of the linear PF equation (\ref{3.11}) in such a system using mathematical
techniques in standard texts \cite{K62}. We first transform Eq. (\ref{3.11})
into the following two second-order elliptic PDEs%
\begin{align}
\nabla^{2}\phi(\mathbf{r})  &  =\frac{1}{\epsilon_{s}}\psi(\mathbf{r})\text{,
}\label{3.15}\\
\nabla^{2}\psi(\mathbf{r})  &  =\frac{-\epsilon_{s}\kappa^{2}}{l_{c}^{2}}%
\phi(\mathbf{r})+\frac{1}{l_{c}^{2}}\psi(\mathbf{r}), \label{3.16}%
\end{align}
by introducing the extra unknown function $\psi(\mathbf{r})$ \cite{L13}, which
is a density-like function as can be seen from Eq. (\ref{2.3}) by setting
$l_{c}=0$. These two equations can be written in matrix form as%
\begin{equation}
\left(
\begin{array}
[c]{c}%
\nabla^{2}\phi(\mathbf{r})\\
\nabla^{2}\psi(\mathbf{r})
\end{array}
\right)  =\left[
\begin{array}
[c]{cc}%
0 & \frac{1}{\epsilon_{s}}\\
\frac{-\epsilon_{s}\kappa^{2}}{l_{c}^{2}} & \frac{1}{l_{c}^{2}}%
\end{array}
\right]  \left(
\begin{array}
[c]{c}%
\phi(\mathbf{r})\\
\psi(\mathbf{r})
\end{array}
\right)  =M\left(
\begin{array}
[c]{c}%
\phi(\mathbf{r})\\
\psi(\mathbf{r})
\end{array}
\right)  . \label{3.17}%
\end{equation}

The characteristic polynomial $g(t)$ of the matrix $M$ is given by%
\[
g(t)=t^{2}-\frac{1}{l_{c}^{2}}t+\frac{\kappa^{2}}{l_{c}^{2}}%
\]
and the distinct roots of $g(t)$ are the eigenvalues%
\begin{equation}
\lambda_{1}=\frac{1-\sqrt{1-4l_{c}^{2}/l_{DPF}^{2}}}{2l_{c}^{2}}\text{,
\ }\lambda_{2}=\frac{1+\sqrt{1-4l_{c}^{2}/l_{DPF}^{2}}}{2l_{c}^{2}}
\label{3.18}%
\end{equation}
of $M$ that can thus be decomposed as%
\[
M=Q\left[
\begin{array}
[c]{cc}%
\lambda_{1} & 0\\
0 & \lambda_{2}%
\end{array}
\right]  Q^{-1}\text{,\ }Q=\left[
\begin{array}
[c]{cc}%
\frac{l_{c}^{2}\lambda_{2}}{\epsilon_{s}\kappa^{2}} & \frac{l_{c}^{2}%
\lambda_{1}}{\epsilon_{s}\kappa^{2}}\\
1 & 1
\end{array}
\right]  \text{, }Q^{-1}=\frac{1}{\lambda_{2}-\lambda_{1}}\left[
\begin{array}
[c]{cc}%
\frac{\epsilon_{s}\kappa^{2}}{l_{c}^{2}} & -\lambda_{1}\\
\frac{-\epsilon_{s}\kappa^{2}}{l_{c}^{2}} & \lambda_{2}%
\end{array}
\right]  .
\]
Denoting%
\begin{equation}
\left(
\begin{array}
[c]{c}%
\phi_{\ast}(\mathbf{r})\\
\psi_{\ast}(\mathbf{r})
\end{array}
\right)  =Q^{-1}\left(
\begin{array}
[c]{c}%
\phi(\mathbf{r})\\
\psi(\mathbf{r})
\end{array}
\right)  \text{,} \label{3.19}%
\end{equation}
Eq. (\ref{3.17}) becomes%
\[
\left(
\begin{array}
[c]{c}%
\nabla^{2}\phi_{\ast}(\mathbf{r})\\
\nabla^{2}\psi_{\ast}(\mathbf{r})
\end{array}
\right)  =\left[
\begin{array}
[c]{cc}%
\lambda_{1} & 0\\
0 & \lambda_{2}%
\end{array}
\right]  \left(
\begin{array}
[c]{c}%
\phi_{\ast}(\mathbf{r})\\
\psi_{\ast}(\mathbf{r})
\end{array}
\right)
\]
which, in spherically symmetric cases, is simplified to%
\begin{align}
\frac{1}{r^{2}}\frac{d}{dr}\left(  r^{2}\frac{d\phi_{\ast}(r)}{dr}\right)   &
=\lambda_{1}\phi_{\ast}(r)\text{,}\label{3.20}\\
\frac{1}{r^{2}}\frac{d}{dr}\left(  r^{2}\frac{d\psi_{\ast}(r)}{dr}\right)   &
=\lambda_{2}\psi_{\ast}(r)\text{,} \label{3.21}%
\end{align}
where $r=\left\vert \mathbf{r}\right\vert $ in $R^{3}$. The general solutions
of these two equations are%
\begin{equation}
\phi_{\ast}(r)=\frac{Ae^{-\sqrt{\lambda_{1}}r}}{r}+\frac{Be^{\sqrt{\lambda
_{1}}r}}{r}\text{, \ }\psi_{\ast}(r)=\frac{Ce^{-\sqrt{\lambda_{2}}r}}{r}%
+\frac{De^{\sqrt{\lambda_{2}}r}}{r}\text{,} \label{3.22}%
\end{equation}
where $A$, $B$, $C$, and $D$ are arbitrary constants. Consequently, the
inverse mapping of Eq. (\ref{3.19}) gives the general solutions of Eqs.
(\ref{3.15}) and (\ref{3.16}) as
\begin{align}
\phi(r)  &  =\frac{l_{c}^{2}\lambda_{2}}{\epsilon_{s}\kappa^{2}}\left(
\frac{Ae^{-\sqrt{\lambda_{1}}r}}{r}+\frac{Be^{\sqrt{\lambda_{1}}r}}{r}\right)
+\frac{l_{c}^{2}\lambda_{1}}{\epsilon_{s}\kappa^{2}}\left(  \frac
{Ce^{-\sqrt{\lambda_{2}}r}}{r}+\frac{De^{\sqrt{\lambda_{2}}r}}{r}\right)
,\label{3.23}\\
\psi(r)  &  =\frac{Ae^{-\sqrt{\lambda_{1}}r}}{r}+\frac{Be^{\sqrt{\lambda_{1}%
}r}}{r}+\frac{Ce^{-\sqrt{\lambda_{2}}r}}{r}+\frac{De^{\sqrt{\lambda_{2}}r}}%
{r}. \label{3.24}%
\end{align}

\subsection{Unique Solution of Linear PF Equation}

The solvation energy -- a fundamental quantity to measure the interaction of a
solute with its solvent -- of an ion in a bulk electrolyte solution can be
calculated by Born models in a geometrically very simple setting \cite{BC00}.
Inspired by Born's work \cite{B20}, Debye and H\"{u}ckel derived a unique
analytical solution of the linear PB equation in spherically symmetric domains
\cite{DH23}. Motivated by Born and Debye-H\"{u}ckel models, a ring-shaped
domain as shown in Fig. 1 was proposed in \cite{LE15a} for solving
\textit{numerically} the nonlinear PF equation (\ref{2.3}), where the system
domain $\overline{\Omega}$ $=$ $\overline{\Omega}_{i}\cup\overline{\Omega
}_{sh}\cup\overline{\Omega}_{s}$ is bounded with the volume $V$, $\Omega_{i}%
$\ is the spherical domain occupied by the solvated ion $i$, $\Omega_{sh}$ is
the hydration shell domain of the ion, $\Omega_{s}$ is the rest of solvent
domain, and O denotes the center (set to the origin $\mathbf{0}$) of the ion.
The radii of $\Omega_{i}$ and the outer boundary of $\Omega_{sh}$ are denoted
by the effective Born (ionic cavity \cite{RH85}) radius $R_{i}^{Born}$ and the
hydration shell radius $R_{i}^{sh}$ (including first and second shells
\cite{RI13}), respectively. \begin{figure}[t]
\centering\includegraphics[scale=0.6]{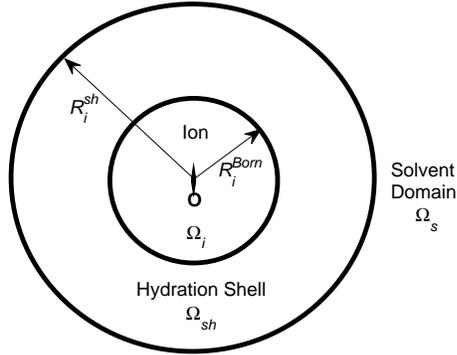}\caption{The model domain
$\Omega$ is partitioned into the ion domain $\Omega_{i}$ (with radius
$R_{i}^{Born}$), the hydration shell domain $\Omega_{sh}$ (with radius
$R_{i}^{sh}$), and the remaining solvent (bounded) domain $\Omega_{s}$.}%
\end{figure}

Debye and H\"{u}ckel \cite{DH23} introduced the activity coefficient
$\gamma_{i}$ of an ion of species $i$ in an electrolyte solution to describe
the deviation of the chemical potential of the ion from ideality ($\gamma
_{i}=1$). The excess chemical potential $\mu_{i}^{ex}=k_{B}T\ln\gamma_{i}$ can
be calculated by \cite{BC00,LE15a}%
\begin{equation}
\mu_{i}^{ex}=\frac{1}{2}q_{i}\phi(\mathbf{0})-\frac{1}{2}q_{i}\phi
^{0}(\mathbf{0})\text{,} \label{3.25}%
\end{equation}
where $\phi(\mathbf{r})$ is a reaction potential \cite{BC00} function and
$\phi^{0}(\mathbf{r})$ is a potential function when the solvent domain
$\Omega_{s}$ does not contain any ions at all with pure water only, i.e., when
the solution is ideal. We seek an algebraic expression of the potential
function $\phi(\mathbf{r})$ by solving \textit{analytically} the following
system of second-order PDEs%
\begin{align}
\epsilon_{s}\nabla^{2}\phi(\mathbf{r})  &  =\psi(\mathbf{r})\text{ in }%
\Omega_{s}\text{,}\label{3.26}\\
\left(  1-l_{c}^{2}\nabla^{2}\right)  \psi(\mathbf{r})  &  =\epsilon_{s}%
\kappa^{2}\phi(\mathbf{r})\text{ in }\Omega_{s}\text{,}\label{3.27}\\
\nabla^{2}\phi(\mathbf{r})  &  =0\text{ in }\Omega_{i}\cup\Omega_{sh}\text{,}
\label{3.28}%
\end{align}
for which the general solutions of (\ref{3.26}) and (\ref{3.27}) are given in
(\ref{3.23}) and (\ref{3.24}), respectively. The boundary and interface
conditions for $\phi(\mathbf{r})$ and $\psi(\mathbf{r})$ are
\begin{align}
\phi(\mathbf{r})  &  =\psi(\mathbf{r})=0\text{ at }\left\vert \mathbf{r}%
\right\vert =\infty,\label{3.29}\\
\psi(\mathbf{r})  &  =\epsilon_{s}\kappa^{2}\phi(\mathbf{r})\text{ on
}\partial\Omega_{sh}\cap\partial\Omega_{s},\label{3.30}\\
\left[  \phi(\mathbf{r})\right]   &  =0\text{ on }\partial\Omega_{i}%
\cup\left(  \partial\Omega_{sh}\cap\partial\Omega_{s}\right)  ,\label{3.31}\\
\left[  \nabla\phi(\mathbf{r})\cdot\mathbf{n}\right]   &  =0\text{ on
}\partial\Omega_{sh}\cap\partial\Omega_{s},\label{3.32}\\
\left[  \epsilon(\mathbf{r})\nabla\phi(\mathbf{r})\cdot\mathbf{n}\right]   &
=\epsilon_{i}\nabla\phi^{\ast}(\mathbf{r})\cdot\mathbf{n}\text{ on }%
\partial\Omega_{i}, \label{3.33}%
\end{align}
where $\partial$ denotes the boundary of a domain, the jump function
$[\phi(\mathbf{r})]=\lim_{\mathbf{r}_{sh}\rightarrow\mathbf{r}}\phi
(\mathbf{r}_{sh})$ $-\lim_{\mathbf{r}_{i}\rightarrow\mathbf{r}}\phi
(\mathbf{r}_{i})$ at $\mathbf{r}$ $\in\partial\Omega_{i}$ with $\mathbf{r}%
_{sh}\in$ $\Omega_{sh}$ and $\mathbf{r}_{i}\in$ $\Omega_{i}$, $\epsilon
(\mathbf{r})=\epsilon_{s}$ in $\Omega_{sh}$ and $\epsilon(\mathbf{r}%
)=\epsilon_{i}$ in $\Omega_{i}$, $\epsilon_{i}=\epsilon_{ion}\epsilon_{0}$,
$\epsilon_{ion}$ is a dielectric constant in $\Omega_{i}$, $\mathbf{n}$ is an
outward normal unit vector at $\mathbf{r}\in$ $\partial\Omega_{i}$, and
$\phi^{\ast}(\mathbf{r})=q_{i}/(4\pi\epsilon_{i}\left\vert \mathbf{r-0}%
\right\vert )$.

The additional Laplace equation (\ref{3.28}) in $\Omega_{i}$ avoids large
errors in a direct approximation of the delta function $\delta(\mathbf{r}%
-\mathbf{0})$ in the singular charge $q_{i}\delta(\mathbf{r}-\mathbf{0})$ of
the solvated ion located at the origin $\mathbf{0}$ by transforming the
singular charge to the Green's function $\phi^{\ast}(\mathbf{r})$ on
$\partial\Omega_{i}$ in Eq. (\ref{3.33}) as an approximate source of the
electric field produced by the solvated ion \cite{CL03,GY07}. From Eqs.
(\ref{2.3}) and (\ref{3.14}), we observe that $\psi(\mathbf{r})=$
$\epsilon_{s}\nabla^{2}\phi(\mathbf{r})=-\rho(\mathbf{r})\approx\epsilon
_{s}\kappa^{2}\phi(\mathbf{r})$ when $l_{c}=0$. Therefore, the interface
condition (\ref{3.30}) simply means that the function $\psi(\mathbf{r})$
satisfying Eq. (\ref{3.27}) in $\Omega_{s}$ is prescribed as a negative charge
density function $\epsilon_{s}\kappa^{2}\phi(\mathbf{r})$ with $l_{c}=0$ on
the boundary $\partial\Omega_{sh}\cap\partial\Omega_{s}$. This interface
condition can be derived from the charge neutrality condition of the entire
system using Gauss's divergence theorem, see Ref. \cite{L13} for the
derivation, where all interface conditions are also presented and treated in
great detail from numerical point of view. Note that, although the entire
electrolyte solution in $\Omega$ is still in bulk condition, the excess
chemical potential of the ion $i$ has been modeled by this PDE system in which
the singular charge of the ion is treated as an external source that generates
the electric potential function $\phi(\mathbf{r})$, i.e., the electrolyte
solution in $\Omega\backslash\Omega_{i}$ is not treated in bulk condition.

The general solution of the Laplace equation (\ref{3.28}) in a spherically
symmetric domain is \cite{K62}%
\[
\phi(r\text{, }\theta)=\sum_{n=1}^{\infty}(A_{n}r^{n}+B_{n}r^{-n-1})P_{n}%
(\cos\theta),
\]
where $\theta$ is the polar angle of a vector $\mathbf{r}$ and $P_{n}%
(\cos\theta)$ are Legendre polynomials. This implies that the solution is
unique and
\begin{equation}
\phi(r)=A_{0}\text{ in }\Omega_{i} \label{3.34}%
\end{equation}
if $\phi(R_{i}^{Born})=A_{0}$, since $r=0$ $\in\Omega_{i}$ and thus $B_{n}=0$
for all $n$, and $A_{n}=0$ for all $n\neq0$ as $r\rightarrow\infty$.
Similarly, the solution is unique and
\begin{equation}
\phi(r)=C_{0}+\frac{D_{0}}{r}\text{ in }\Omega_{sh} \label{3.35}%
\end{equation}
if $\phi(R_{i}^{Born})=C_{0}+\frac{D_{0}}{R_{i}^{Born}}$ and $\phi(R_{i}%
^{sh})=C_{0}+\frac{D_{0}}{R_{i}^{sh}}$. We need 7 conditions to uniquely
determine the 7 unknowns $A$, $B$, $C$, $D$, $A_{0}$, $C_{0}$, and $D_{0}$,
and to prove that $\phi(R_{i}^{Born})$ and $\phi(R_{i}^{sh})$ are constants.

\textit{Conds. 1 and 2.} By $\phi(\mathbf{r})=0$ and $\psi(\mathbf{r})=0$ in
(\ref{3.29}) as $r\rightarrow\infty$, (\ref{3.23}), and (\ref{3.24}), we
obtain $B=D=0$ and hence%
\begin{align*}
\phi(r)  &  =\frac{Al_{c}^{2}\lambda_{2}}{\epsilon_{s}\kappa^{2}}%
\frac{e^{-\sqrt{\lambda_{1}}r}}{r}+\frac{Cl_{c}^{2}\lambda_{1}}{\epsilon
_{s}\kappa^{2}}\frac{e^{-\sqrt{\lambda_{2}}r}}{r}\text{ in }\Omega_{s},\\
\psi(r)  &  =\frac{Ae^{-\sqrt{\lambda_{1}}r}}{r}+\frac{Ce^{-\sqrt{\lambda_{2}%
}r}}{r}\text{ in }\Omega_{s}.
\end{align*}

\textit{Cond. 3.} $\left[  \phi(\mathbf{r})\right]  =0$ on $\partial
\Omega_{sh}\cap\partial\Omega_{s}$ in (\ref{3.31}) and (\ref{3.35}), we have
\[
\phi(R_{i}^{sh})=\frac{Al_{c}^{2}\lambda_{2}}{\epsilon_{s}\kappa^{2}}%
\frac{e^{-\sqrt{\lambda_{1}}R_{i}^{sh}}}{R_{i}^{sh}}+\frac{Cl_{c}^{2}%
\lambda_{1}}{\epsilon_{s}\kappa^{2}}\frac{e^{-\sqrt{\lambda_{2}}R_{i}^{sh}}%
}{R_{i}^{sh}}=C_{0}+\frac{D_{0}}{R_{i}^{sh}}\text{,}%
\]
which implies that $\phi(R_{i}^{sh})$ is constant and
\[
A\lambda_{2}e^{-\sqrt{\lambda_{1}}R_{i}^{sh}}+C\lambda_{1}e^{-\sqrt
{\lambda_{2}}R_{i}^{sh}}=\frac{\epsilon_{s}\kappa^{2}}{l_{c}^{2}}\left(
C_{0}R_{i}^{sh}+D_{0}\right)  \text{.}%
\]

\textit{Cond. 4.} By (\ref{3.30}), we have%
\begin{align*}
\psi(R_{i}^{sh})  &  =\frac{Ae^{-\sqrt{\lambda_{1}}R_{i}^{sh}}}{R_{i}^{sh}%
}+\frac{Ce^{-\sqrt{\lambda_{2}}R_{i}^{sh}}}{R_{i}^{sh}}=\epsilon_{s}\kappa
^{2}\phi(R_{i}^{sh})\\
&  =\frac{Al_{c}^{2}\lambda_{2}e^{-\sqrt{\lambda_{1}}R_{i}^{sh}}}{R_{i}^{sh}%
}+\frac{Cl_{c}^{2}\lambda_{1}e^{-\sqrt{\lambda_{2}}R_{i}^{sh}}}{R_{i}^{sh}},
\end{align*}%
\[
Ae^{-\sqrt{\lambda_{1}}R_{i}^{sh}}\left(  l_{c}^{2}\lambda_{2}-1\right)
+Ce^{-\sqrt{\lambda_{2}}R_{i}^{sh}}\left(  l_{c}^{2}\lambda_{1}-1\right)  =0.
\]

\textit{Cond. 5.} By (\ref{3.32}), we have
\begin{align*}
\left[  \frac{\partial\phi(\mathbf{r})}{\partial\mathbf{n}}\right]   &
=\lim_{r\mathbf{\rightarrow}R_{i}^{sh}}\frac{d\left(  \frac{Al_{c}^{2}%
\lambda_{2}}{\epsilon_{s}\kappa^{2}}\frac{e^{-\sqrt{\lambda_{1}}r}}{r}%
+\frac{Cl_{c}^{2}\lambda_{1}}{\epsilon_{s}\kappa^{2}}\frac{e^{-\sqrt
{\lambda_{2}}r}}{r}\right)  }{dr}-\lim_{r\mathbf{\rightarrow}R_{i}^{sh}}%
\frac{d\left(  C_{0}+\frac{D_{0}}{r}\right)  }{dr}\\
&  =\lim_{r\mathbf{\rightarrow}R_{i}^{sh}}\left[  \frac{-Al_{c}^{2}\lambda
_{2}\left(  \sqrt{\lambda_{1}}r+1\right)  }{\epsilon_{s}\kappa^{2}}%
\frac{e^{-\sqrt{\lambda_{1}}r}}{r^{2}}\right. \\
&  -\left.  \frac{Cl_{c}^{2}\lambda_{1}\left(  \sqrt{\lambda_{2}}r+1\right)
}{\epsilon_{s}\kappa^{2}}\frac{e^{-\sqrt{\lambda_{2}}r}}{r^{2}}+\frac{D_{0}%
}{r^{2}}\right] \\
&  =\frac{1}{\left(  R_{i}^{sh}\right)  ^{2}}\left[  D_{0}-\frac{Al_{c}%
^{2}\lambda_{2}\left(  \sqrt{\lambda_{1}}R_{i}^{sh}+1\right)  e^{-\sqrt
{\lambda_{1}}R_{i}^{sh}}}{\epsilon_{s}\kappa^{2}}\right. \\
&  -\left.  \frac{Cl_{c}^{2}\lambda_{1}\left(  \sqrt{\lambda_{2}}R_{i}%
^{sh}+1\right)  e^{-\sqrt{\lambda_{2}}R_{i}^{sh}}}{\epsilon_{s}\kappa^{2}%
}\right]
\end{align*}%
\[
A\lambda_{2}\left(  \sqrt{\lambda_{1}}R_{i}^{sh}+1\right)  e^{-\sqrt
{\lambda_{1}}R_{i}^{sh}}+C\lambda_{1}\left(  \sqrt{\lambda_{2}}R_{i}%
^{sh}+1\right)  e^{-\sqrt{\lambda_{2}}R_{i}^{sh}}=\frac{D_{0}\epsilon
_{s}\kappa^{2}}{l_{c}^{2}}.
\]

\textit{Cond. 6.} By $\left[  \phi(\mathbf{r})\right]  =0$ on $\partial
\Omega_{i}$ in (\ref{3.31}), we have%
\[
\phi(R_{i}^{Born})=C_{0}+\frac{D_{0}}{R_{i}^{Born}}=A_{0}\text{,}%
\]
which implies that $\phi(R_{i}^{Born})$ is constant.

\textit{Cond. 7.} By (\ref{3.33}), we have%
\begin{align*}
\left[  \epsilon(\mathbf{r})\frac{\partial\phi(\mathbf{r})}{\partial
\mathbf{n}}\right]   &  =\epsilon_{s}\lim_{r\mathbf{\rightarrow}R_{i}^{Born}%
}\frac{d\left(  C_{0}+\frac{D_{0}}{r}\right)  }{dr}-\epsilon_{i}%
\lim_{r\mathbf{\rightarrow}R_{i}^{Born}}\frac{dA_{0}}{dr}=-\frac{\epsilon
_{s}D_{0}}{\left(  R_{i}^{Born}\right)  ^{2}}\\
&  =\epsilon_{i}\nabla\phi^{\ast}(\mathbf{r})\cdot\mathbf{n}\text{ }%
=\frac{q_{i}}{4\pi}\lim_{r\mathbf{\rightarrow}R_{i}^{Born}}\frac{d}{dr}%
\frac{1}{r}=-\frac{q_{i}}{4\pi\left(  R_{i}^{Born}\right)  ^{2}}.
\end{align*}
Therefore, we find%
\begin{align*}
A  &  =\frac{q_{i}\kappa^{2}}{4\pi l_{c}^{2}}\left[  \frac{e^{\sqrt
{\lambda_{1}}R_{i}^{sh}}\left(  l_{c}^{2}\lambda_{1}-1\right)  }{\lambda
_{2}\left(  l_{c}^{2}\lambda_{1}-1\right)  \left(  \sqrt{\lambda_{1}}%
R_{i}^{sh}+1\right)  -\lambda_{1}\left(  l_{c}^{2}\lambda_{2}-1\right)
\left(  \sqrt{\lambda_{2}}R_{i}^{sh}+1\right)  }\right]  ,\\
C  &  =\frac{q_{i}\kappa^{2}}{4\pi l_{c}^{2}}\left[  \frac{-e^{\sqrt
{\lambda_{2}}R_{i}^{sh}}\left(  l_{c}^{2}\lambda_{2}-1\right)  }{\lambda
_{2}\left(  l_{c}^{2}\lambda_{1}-1\right)  \left(  \sqrt{\lambda_{1}}%
R_{i}^{sh}+1\right)  -\lambda_{1}\left(  l_{c}^{2}\lambda_{2}-1\right)
\left(  \sqrt{\lambda_{2}}R_{i}^{sh}+1\right)  }\right]  ,\\
B  &  =D=0,\\
C_{0}  &  =\frac{q_{i}}{4\pi\epsilon_{s}R_{i}^{sh}}\left[  \frac{\lambda
_{2}\left(  l_{c}^{2}\lambda_{1}-1\right)  -\lambda_{1}\left(  l_{c}%
^{2}\lambda_{1}-1\right)  }{\lambda_{2}\left(  l_{c}^{2}\lambda_{1}-1\right)
\left(  \sqrt{\lambda_{1}}R_{i}^{sh}+1\right)  -\lambda_{1}\left(  l_{c}%
^{2}\lambda_{2}-1\right)  \left(  \sqrt{\lambda_{2}}R_{i}^{sh}+1\right)
}-1\right]  ,\\
D_{0}  &  =\frac{q_{i}}{4\pi\epsilon_{s}},\\
A_{0}  &  =C_{0}+\frac{D_{0}}{R_{i}^{Born}}.
\end{align*}
Since%
\begin{align*}
\lambda_{1}+\lambda_{2}  &  =\frac{1}{l_{c}^{2}}\text{,}\\
\lambda_{2}\left(  l_{c}^{2}\lambda_{1}-1\right)   &  =\lambda_{2}\left(
\frac{\lambda_{1}}{\text{ }\lambda_{1}+\lambda_{2}}-1\right)  =\frac
{-\lambda_{2}^{2}}{\lambda_{1}+\lambda_{2}}=-l_{c}^{2}\lambda_{2}^{2}%
\text{,}\\
\lambda_{1}\left(  l_{c}^{2}\lambda_{2}-1\right)   &  =-l_{c}^{2}\lambda
_{1}^{2}\text{,}%
\end{align*}
we introduce the symbol $\Theta$ for $C_{0}$ such that
\begin{align}
\Theta &  =\frac{\lambda_{2}\left(  l_{c}^{2}\lambda_{1}-1\right)
-\lambda_{1}\left(  l_{c}^{2}\lambda_{1}-1\right)  }{\lambda_{2}\left(
l_{c}^{2}\lambda_{1}-1\right)  \left(  \sqrt{\lambda_{1}}R_{i}^{sh}+1\right)
-\lambda_{1}\left(  l_{c}^{2}\lambda_{2}-1\right)  \left(  \sqrt{\lambda_{2}%
}R_{i}^{sh}+1\right)  }\nonumber\\
&  =\frac{-l_{c}^{2}\lambda_{2}^{2}+l_{c}^{2}\lambda_{1}^{2}}{-l_{c}%
^{2}\lambda_{2}^{2}\left(  \sqrt{\lambda_{1}}R_{i}^{sh}+1\right)  +l_{c}%
^{2}\lambda_{1}^{2}\left(  \sqrt{\lambda_{2}}R_{i}^{sh}+1\right)  }\nonumber\\
&  =\frac{\lambda_{1}^{2}-\lambda_{2}^{2}}{\lambda_{1}^{2}\left(
\sqrt{\lambda_{2}}R_{i}^{sh}+1\right)  -\lambda_{2}^{2}\left(  \sqrt
{\lambda_{1}}R_{i}^{sh}+1\right)  }\nonumber\\
&  =\frac{\lambda_{1}-\lambda_{2}}{l_{c}^{2}\lambda_{1}^{2}\left(
\sqrt{\lambda_{2}}R_{i}^{sh}+1\right)  -l_{c}^{2}\lambda_{2}^{2}\left(
\sqrt{\lambda_{1}}R_{i}^{sh}+1\right)  }. \label{3.36}%
\end{align}
We summarize our analysis as the following main result of the current study.

\textbf{Theorem 3.2.} For a binary aqueous electrolytic solution in a
spherically symmetric domain as shown in Fig. 1, the linear Poisson-Fermi
model system (\ref{3.26}) -- (\ref{3.33}) has the unique potential function%
\begin{equation}
\phi^{PF}(r)=\left\{
\begin{array}
[c]{l}%
\frac{q_{i}}{4\pi\epsilon_{s}R_{i}^{Born}}+\frac{q_{i}}{4\pi\epsilon_{s}%
R_{i}^{sh}}\left(  \Theta-1\right)  \text{ in }\Omega_{i}\\
\frac{q_{i}}{4\pi\epsilon_{s}r}+\frac{q_{i}}{4\pi\epsilon_{s}R_{i}^{sh}%
}\left(  \Theta-1\right)  \text{ in }\Omega_{sh}\\
\frac{q_{i}}{4\pi\epsilon_{s}r}\left[  \frac{\lambda_{1}^{2}e^{-\sqrt
{\lambda_{2}}\left(  r-R_{i}^{sh}\right)  }-\lambda_{2}^{2}e^{-\sqrt
{\lambda_{1}}\left(  r-R_{i}^{sh}\right)  }}{\lambda_{1}^{2}\left(
\sqrt{\lambda_{2}}R_{i}^{sh}+1\right)  -\lambda_{2}^{2}\left(  \sqrt
{\lambda_{1}}R_{i}^{sh}+1\right)  }\right]  \text{ in }\Omega_{s}.
\end{array}
\right.  \label{3.37}%
\end{equation}

\textbf{Remark 3.3.} Note that $\lim_{l_{c}\rightarrow0}\lambda_{1}%
=1/l_{DPF}^{2}$ (correlation effect is ignored), $\lim_{l_{c}\rightarrow
0}\lambda_{2}=\infty$, and $\lim_{l_{c}\rightarrow0}\Theta=\lim_{C_{1}%
^{B}\rightarrow0}\Theta=\lim_{l_{DPF}\rightarrow\infty}\Theta=1$ (correlation
is ignored and electrolyte is infinite dilute). The linearized PF potential
$\phi^{PF}(r)$ reduces to the linearized PB potential $\phi^{PB}%
(r)=q_{i}e^{-r/l_{D}}/(4\pi\epsilon_{s}r)$ as in standard texts (e.g. Eq.
(7.46) in \cite{LM03}) by taking $\lim_{l_{c}\rightarrow0}\phi^{PF}(r)$ with
$v_{j}=0$ for all $j$ (steric effect is ignored), $R_{i}^{sh}=0$, and $r>0$.

\subsection{Generalized Debye-H\"{u}ckel Equation}

As discussed in \cite{VB15}, the solvation free energy of an ion $i$ should
vary with salt concentrations, i.e., the Born energy
\begin{equation}
-\frac{q_{i}^{2}}{8\pi\epsilon_{0}R_{i}^{0}}\left(  1-\frac{1}{\epsilon_{w}%
}\right)  \label{3.37a}%
\end{equation}
in pure water (i.e. $C_{i}^{B}=0$) with a constant Born radius $R_{i}^{0}$
should be modified to depend on $C_{i}^{B}\geq0$. Equivalently, the effective
Born radius $R_{i}^{Born}$ of the electrolyte solution in Fig. 1 varies with
$C_{i}^{B}$ and can be modified from $R_{i}^{0}$ by a simple formula
\cite{LE15a}
\begin{equation}
R_{i}^{Born}(C_{i}^{B})=\theta(C_{i}^{B})R_{i}^{0}\text{, \ \ }\theta
(C_{i}^{B})=1+\alpha_{1}^{i}\left(  \overline{C}_{i}^{B}\right)  ^{1/2}%
+\alpha_{2}^{i}\overline{C}_{i}^{B}+\alpha_{3}^{i}\left(  \overline{C}_{i}%
^{B}\right)  ^{3/2}\text{,} \label{3.38}%
\end{equation}
where $\overline{C}_{i}^{B}=$ $C_{i}^{B}$/M is a dimensionless bulk
concentration, M is molarity (molar concentration), and $\alpha_{1}^{i}$,
$\alpha_{2}^{i}$, and $\alpha_{3}^{i}$ are adjustable parameters for modifying
the experimental Born radius $R_{i}^{0}$ to fit experimental activity
coefficients $\gamma_{i}$ that change with the bulk concentration $C_{i}^{B}$
of the ion. The Born radii $R_{i}^{0}$ given below are cited from \cite{VB15},
which are computed from the experimental hydration Helmholtz free energies of
these ions given in \cite{F04}. The three parameters in (\ref{3.38}) have
physical or mathematical meanings unlike numerous parameters in the Pitzer
model \cite{F10,RK15,V11}. The first parameter $\alpha_{1}^{i}$ is an
adjustment of $R_{i}^{0}$ that accounts for the real thickness of the ionic
atmosphere (Debye length), which is proportional to the square root of the
ionic strength $I=\frac{1}{2}\sum_{i}C_{i}^{B}z_{i}^{2}$ in the DH theory
\cite{LM03}. The second $\alpha_{2}^{i}$ and third $\alpha_{3}^{i}$ parameters
are adjustments in the next orders of approximation beyond the DH treatment of
ionic atmosphere \cite{LE15a}.

The potential value $\phi^{0}(\mathbf{0})=\lim_{C_{1}^{B}\rightarrow0}%
\phi^{PF}(\mathbf{0})=$ $q_{i}/\left(  4\pi\epsilon_{s}R_{i}^{0}\right)  $ by
$\lim_{C_{1}^{B}\rightarrow0}\Theta=1$ and $\lim_{C_{1}^{B}\rightarrow0}%
R_{i}^{Born}(C_{i}^{B})=R_{i}^{0}$. From (\ref{3.25}) and (\ref{3.37}), we
thus obtain a generalized activity coefficient $\gamma_{i}^{DHPF}$ as%
\begin{equation}
\ln\gamma_{i}^{DHPF}=\frac{q_{i}^{2}}{8\pi\epsilon_{s}k_{B}T}\left(  \frac
{1}{R_{i}^{Born}(C_{i}^{B})}-\frac{1}{R_{i}^{0}}+\frac{\Theta-1}{R_{i}^{sh}%
}\right)  . \label{3.39}%
\end{equation}
Since the steric potential $S^{\text{trc}}(\mathbf{r})$ in (\ref{2.2}) takes
particle volumes and voids into account, the shell volume $V_{sh}$ of the
shell domain $\Omega_{sh}$ can be determined by the steric potential
\begin{equation}
S_{sh}^{\text{trc}}=\frac{v_{0}}{v_{w}}\ln\frac{O_{i}^{w}}{V_{sh}C_{K+1}^{B}%
}=\ln\frac{V_{sh}-v_{w}O_{i}^{w}}{V_{sh}\Gamma^{B}} \label{3.39a}%
\end{equation}
\cite{LE15a}, where the occupant (coordination) number $O_{i}^{w}$ of water
molecules is given by experimental data \cite{RI13}. The shell radius
$R_{i}^{sh}$ is thus determined and depends not only on $O_{i}^{w}$ but also
on the bulk void fraction $\Gamma^{B}$, namely, \textit{on all salt and water
bulk concentrations} ($C_{k}^{B}$).

\textbf{Remark 3.4.} The generalized activity coefficient $\gamma_{i}^{DHPF} $
reduces to the classical $\gamma_{i}^{DH}$ proposed by Debye and H\"{u}ckel in
1923 \cite{DH23}, namely,
\begin{equation}
\ln\gamma_{i}^{DH}=\frac{-q_{i}^{2}}{8\pi\epsilon_{s}k_{B}T(R_{i}+l_{D})}
\label{3.40}%
\end{equation}
provided that $R_{i}^{Born}(C_{i}^{B})=R_{i}^{0}$ (without considering the
Born energy effect), $R_{i}^{sh}=R_{i}$ (an effective ionic radius (parameter)
\cite{DH23}), $l_{DPF}=l_{D}$ (no steric effect), and $l_{c}=0$ (no
correlation effect). The reduction is shown by taking the limit of the last
term in (\ref{3.39}) as $l_{c}\rightarrow0$, i.e.,
\[
\lim_{l_{c}\rightarrow0}\frac{\Theta-1}{R_{i}^{sh}}=\frac{-1}{R_{i}+l_{D}}%
\]
since%
\begin{align*}
\Theta &  =\frac{\lambda_{1}^{2}-\lambda_{2}^{2}}{\lambda_{1}^{2}\left(
\sqrt{\lambda_{2}}R_{i}^{sh}+1\right)  -\lambda_{2}^{2}\left(  \sqrt
{\lambda_{1}}R_{i}^{sh}+1\right)  }\\
\Theta-1  &  =\frac{-\left(  \lambda_{1}^{2}\sqrt{\lambda_{2}}R_{i}%
^{sh}-\lambda_{2}^{2}\sqrt{\lambda_{1}}R_{i}^{sh}\right)  }{\lambda_{1}%
^{2}\left(  \sqrt{\lambda_{2}}R_{i}^{sh}+1\right)  -\lambda_{2}^{2}\left(
\sqrt{\lambda_{1}}R_{i}^{sh}+1\right)  }\\
\frac{\Theta-1}{R_{i}^{sh}}  &  =\frac{-\left(  \lambda_{1}^{2}\sqrt
{\lambda_{2}}-\lambda_{2}^{2}\sqrt{\lambda_{1}}\right)  }{R_{i}^{sh}\left(
\lambda_{1}^{2}\sqrt{\lambda_{2}}-\lambda_{2}^{2}\sqrt{\lambda_{1}}\right)
+\left(  \lambda_{1}^{2}-\lambda_{2}^{2}\right)  }=\frac{-1}{R_{i}^{sh}+G}\\
\text{\ }G  &  =\frac{\lambda_{1}^{2}-\lambda_{2}^{2}}{\lambda_{1}^{2}%
\sqrt{\lambda_{2}}-\lambda_{2}^{2}\sqrt{\lambda_{1}}}=\frac{\lambda_{1}%
^{2}/\lambda_{2}^{2}-1}{\lambda_{1}^{2}\sqrt{\lambda_{2}}/\lambda_{2}%
^{2}-\sqrt{\lambda_{1}}}\\
\lim_{l_{c}\rightarrow0}G  &  =\lim_{l_{c}\rightarrow0}\frac{1}{\sqrt
{\lambda_{1}}}=l_{D}\text{, }\left(  \lim_{l_{c}\rightarrow0}\lambda_{1}%
=\frac{1}{l_{D}^{2}}\text{, }\lim_{l_{c}\rightarrow0}\lambda_{2}%
=\infty\right)  \text{.}%
\end{align*}
H\"{u}ckel soon realized that the DH formula (\ref{3.40}) failed to fit
experimental data at high ionic strengths and modified it in 1925 \cite{H25}
by adding a linear term in $C_{i}^{B}$ with an extra parameter $\eta_{1}^{i} $
to become
\begin{equation}
\ln\gamma_{i}^{DHB}=\ln\gamma_{i}^{DH}+\eta_{1}^{i}C_{i}^{B}, \label{3.41}%
\end{equation}
where the linear term is an approximation of the Born solvation energy%
\begin{equation}
\frac{q_{i}^{2}}{8\pi\epsilon_{0}R_{i}^{0}}\left(  \frac{1}{\epsilon_{w}%
}-\frac{1}{\epsilon}\right)  \label{3.42}%
\end{equation}
as the permittivity varies from $\epsilon_{w}\epsilon_{0}$ in pure water to
$\epsilon\epsilon_{0}$ in electrolyte solutions, where the dielectric constant
$\epsilon$ is unknown and changes with $C_{i}^{B}$, i.e., $\epsilon
=\epsilon(C_{i}^{B})$ a function of salt concentrations. Consequently, a
variety of extended DH models $\gamma_{i}^{DHBx}$ \cite{RK15} in the form
similar to%
\begin{equation}
\ln\gamma_{i}^{DHBx}=\ln\gamma_{i}^{DH}+\sum_{k=1}^{n}\eta_{k}^{i}\left(
C_{i}^{B}\right)  ^{k} \label{3.43}%
\end{equation}
have been proposed in the literature to express other thermodynamic properties
such as temperature and pressure by a power expansion of $C_{i}^{B}$ with more
and more adjustable parameters $\eta_{k}^{i}$ that can increase
combinatorially with various composition (various $i$), temperature, and
pressure to a frustrating amount as mentioned above. Note that $\eta_{k}^{i}$
may also depend on ionic strength $I$ in a complicated way, see e.g. Eq. (2)
in \cite{RK15}. Many expressions of those parameters are rather long and
tedious and do not have clear physical meaning \cite{F10,RK15,V11}. Moreover,
it has been reported in \cite{FT96} that no improvement is found for the
extended DH model (\ref{3.41}) by changing the constants in the approximation
of $\epsilon(C_{i}^{B})$ to reflect changes in solvent permittivity. This
means that changing only $\eta_{1}^{i}$ may not improve the model to fit
experimental data unless more adjustable parameters are introduced to model
the Born energy more accurately as proposed in \cite{SL15}. \ 

The $R_{i}^{Born}$ term in (\ref{3.39}) differs significantly from the last
term in (\ref{3.43}) as they are inverse of each other in terms of $C_{i}^{B}
$ and parameters. Therefore, the generalized $\gamma_{i}^{DHPF}$ is not a
$\gamma_{i}^{DHBx}$ for which the empirical nature of extended DH models
requires a great deal of efforts to extract parameters (without physical
hints) from existent thermodynamic databases by regression analysis
\cite{RK15,V11,VB07}.

\textbf{Remark 3.5.} Same as the classical $\gamma_{i}^{DH}$, the generalized
activity coefficient also satisfies the DH limiting law \cite{LM03}, i.e.,
$\gamma_{i}^{DHPF}=\gamma_{i}^{DH}=1$ as $C_{i}^{B}\rightarrow0$ for infinite
dilute (ideal) solutions. The DH limiting law is useful for calculating the
activity coefficient of an ion in very dilute solutions to compare with
experimental measurements that are especially important for highly charged
electrolytes \cite{F18}.

The formula (\ref{3.39}) shows that the principal determinant of ionic
activity is the concentration-dependent Born radius $R_{i}^{Born}(C_{i}^{B})$
since $\gamma_{i}^{DHPF}$ is very sensitive to $R_{i}^{Born}(C_{i}^{B})$ that
is an atomic distance from the singular charge $q_{i}\delta(\mathbf{r}%
-\mathbf{0})$ of the ion, which is infinite at $\mathbf{0}$ and thus
critically affects $\gamma_{i}^{DHPF}$. The secondary determinant is the
hydration shell radius $R_{i}^{sh}$ that lumps short-range ion-water
interactions into a single physical length. The last part of ionic activity is
extracted to the symbol $\Theta$ that accounts for ion-ion correlations
($l_{c}$) and long-range electrostatics ($l_{DPF}$) via the eigenvalues
$\lambda_{1}$ and $\lambda_{2}$ in (\ref{3.18}) and (\ref{3.36}).

\section{Conclusion}

A generalized Debye-H\"{u}ckel equation has been derived and analyzed from the
Poisson-Fermi theory that accounts for the steric, correlation, and
polarization effects of ions and water in aqueous electrolyte solutions at
variable composition, concentration, temperature, and pressure. A generalized
Debye length is proposed to include the size effect of ions and water. The new
equation and length have been shown to reduce to their classical counterparts
when these three effects are ignored. We have also shown that the generalized
DH model is not an extended Debye-H\"{u}ckel model since their approximations
of the Born solvation energy are inversely different in salt concentration.

\textbf{Acknowledgement.} This work was supported by the Ministry of Science
and Technology, Taiwan (No. MOST 105-2115-M-007-016-MY2 to J.L.L.).


\begin{thebibliography}{99}                                                                                               %


\bibitem {BK09}M. Z. Bazant, M. S. Kilic, B. D. Storey, and A. Ajdari, Towards
an understanding of induced-charge electrokinetics at large applied voltages
in concentrated solutions, Adv. Coll. Interf. Sci. \textbf{152}, 48-88 (2009).

\bibitem {BS11}M. Z. Bazant, B. D. Storey, and A. A. Kornyshev, Double layer
in ionic liquids: Overscreening versus crowding, Phys. Rev. Lett.
\textbf{106}, 046102 (2011).

\bibitem {BC00}D. Bashford and D. A. Case, Generalized Born models of
macromolecular solvation effects, Annu. Rev. Phys. Chem. \textbf{51}, 129-152 (2000).

\bibitem {B00}H. M. Berman et al., The protein data bank, Nucleic Acids Res.
\textbf{28}, 235-242 (2000).

\bibitem {B20}M. Born, Volumen und hydratationsw\"{a}rme der ionen, Z. Phys.
\textbf{1}, 45-48 (1920).

\bibitem {C13}D. L. Chapman, A contribution to the theory of
electrocapillarity, Phil. Mag. \textbf{25}, 475-481 (1913).

\bibitem {CL03}I-L. Chern, J.-G. Liu, and W.-C. Wang, Accurate evaluation of
electrostatics for macromolecules in solution, Methods Appl. Anal.
\textbf{10}, 309-328 (2003).

\bibitem {DH23}P. Debye and E. H\"{u}ckel, Zur Theorie der Elektrolyte. I.
Gefrierpunktserniedrigung und verwandte Erscheinunge (The theory of
electrolytes. I. Lowering of freezing point and related phenomena), Phys.
Zeitschr. \textbf{24}, 185-206 (1923).

\bibitem {E13}B. Eisenberg, Interacting ions in Biophysics: Real is not ideal,
Biophys. J. \textbf{104}, 1849-1866 (2013).

\bibitem {F04}W. R. Fawcett, \textit{Liquids, Solutions, and Interfaces: From
Classical Macroscopic Descriptions to Modern Microscopic Details} (Oxford
University Press, New York, 2004).

\bibitem {FT96}W. R. Fawcett and A. C. Tikanen, Role of solvent permittivity
in estimation of electrolyte activity coefficients on the basis of the mean
spherical approximation, J. Chem. Phys. \textbf{100}, 4251-4255 (1996).

\bibitem {F10}D. Fraenkel, Simplified electrostatic model for the
thermodynamic excess potentials of binary strong electrolyte solutions with
size-dissimilar ions, Mol. Phys. \textbf{108}, 1435 (2010).

\bibitem {F18}D. Fraenkel, Negative deviations from the Debye-H\"{u}ckel
limiting law for high-charge polyvalent electrolytes: Are they real?, J. Chem.
Theory Comput. \textbf{14}, 2609-2620 (2018).

\bibitem {GY07}W. Geng, S. Yu, and G. Wei, Treatment of charge singularities
in implicit solvent models, J. Chem. Phys. \textbf{127}, 114106 (2007).

\bibitem {G10}M. Gouy, Sur la constitution de la charge electrique a la
surface d'un electrolyte (Constitution of the electric charge at the surface
of an electrolyte), J. Phys. \textbf{9}, 457-468 (1910).

\bibitem {GM09}L. Gross, et al., The chemical structure of a molecule resolved
by atomic force microscopy, Science \textbf{325}, 1110-1114 (2009).

\bibitem {H01}B. Hille, \textit{Ionic Channels of Excitable Membranes}
(Sinauer Associates Inc., Sunderland, MA, 2001).

\bibitem {H25}E. H\"{u}ckel, Zur Theorie konzentrierterer w\"{a}sseriger
L\"{o}sungen starker Elektrolyte, Phys. Z. \textbf{26}, 93-147 (1925).

\bibitem {KF09}G. M. Kontogeorgis and G. K. Folas, \textit{Thermodynamic
Models for Industrial Applications: From Classical and Advanced Mixing Rules
to Association Theories} (John Wiley \& Sons, 2009).

\bibitem {KM18}G. M. Kontogeorgis, B. Maribo-Mogensen, and K. Thomsen, The
Debye-H\"{u}ckel theory and its importance in modeling electrolyte solutions,
Fluid Phase Equil. \textbf{462}, 130-152 (2018).

\bibitem {K62}E. Kreyszig, \textit{Advanced Engineering Mathematics} (Wiley,
1st ed. 1962, 10th ed. 2011).

\bibitem {K10}W. Kunz, \textit{Specific Ion Effects} (World Scientific,
Singapore 2010).

\bibitem {LM03}K. J. Laidler, J. H. Meiser, and B. C. Sanctuary,
\textit{Physical Chemistry} (Houghton Mifflin Co., Boston, 2003).

\bibitem {LJ08}G. Lebon, D. Jou, and J. Casas-V\'{a}zquez,
\textit{Understanding Non-equilibrium Thermodynamics: Foundations,
Applications, Frontiers} (Springer, 2008).

\bibitem {LF96}B. P. Lee and M. E. Fisher, Density fluctuations in an
electrolyte from generalized Debye-Hueckel theory, Phys. Rev. Lett.
\textbf{76}, 2906 (1996).

\bibitem {L13}J.-L. Liu, Numerical methods for the Poisson-Fermi equation in
electrolytes, J. Comput. Phys. \textbf{247}, 88-99 (2013).

\bibitem {LE13}J.-L. Liu and B. Eisenberg, Correlated ions in a calcium
channel model: a Poisson-Fermi theory, J. Phys. Chem. B \textbf{117},
12051-12058 (2013).

\bibitem {LE14}J.-L. Liu and B. Eisenberg, Poisson-Nernst-Planck-Fermi theory
for modeling biological ion channels, J. Chem. Phys. \textbf{141}, 22D532 (2014).

\bibitem {LE14a}J.-L. Liu and B. Eisenberg, Analytical models of calcium
binding in a calcium channel, J. Chem. Phys. \textbf{141}, 075102 (2014).

\bibitem {LE15}J.-L. Liu and B. Eisenberg, Numerical methods for a
Poisson-Nernst-Planck-Fermi model of biological ion channels, Phys. Rev. E
\textbf{92}, 012711 (2015).

\bibitem {LE15a}J.-L. Liu and B. Eisenberg, Poisson-Fermi model of single ion
activities in aqueous solutions, Chem. Phys. Lett. \textbf{637}, 1-6 (2015).

\bibitem {LE18}J.-L. Liu and B. Eisenberg, Poisson-Fermi modeling of ion
activities in aqueous single and mixed electrolyte solutions at variable
temperature, J. Chem. Phys. \textbf{148}, 054501 (2018).

\bibitem {LH16}J.-L. Liu, H.-j. Hsieh, and B. Eisenberg, Poisson-Fermi
modeling of the ion exchange mechanism of the sodium/calcium exchanger, J.
Phys. Chem. B \textbf{120}, 2658-2669 (2016).

\bibitem {LX17}J.-L. Liu, D. Xie, and B. Eisenberg, Poisson-Fermi formulation
of nonlocal electrostatics in electrolyte solutions, Mol. Based Math. Biol.
\textbf{5}, 116-124 (2017).

\bibitem {N91}J. Newman, \textit{Electrochemical Systems} (Prentice-Hall, NJ, 1991).

\bibitem {MP11}J. M\"{a}hler and I. Persson, A study of the hydration of the
alkali metal ions in aqueous solution, Inorg. Chem. \textbf{51}, 425 (2011).

\bibitem {P95}K. S. Pitzer, \textit{Thermodynamics} (McGraw Hill, New York, 1995).

\bibitem {RH85}A. A. Rashin and B. Honig, Reevaluation of the Born model of
ion hydration, J. Phys. Chem. \textbf{89}, 5588-5593 (1985).

\bibitem {RS59}R. Robinson and R. Stokes, \textit{Electrolyte Solutions}
(Butterworths Scientific Publications, London, 1959); (Dover Publications, New
York, 2002).

\bibitem {RK15}D. Rowland, E. K\"{o}nigsberger, G. Hefter, and P. M. May,
Aqueous electrolyte solution modelling: Some limitations of the Pitzer
equations, Appl. Geochem. \textbf{55}, 170 (2015).

\bibitem {RI13}W. W. Rudolph and G. Irmer, Hydration of the calcium(II) ion in
an aqueous solution of common anions (ClO$_{4}^{-}$, Cl$^{-}$, Br$^{-}$, and
NO$_{3}^{-}$), Dalton Trans. \textbf{42}, 3919 (2013).

\bibitem {S06}C. D. Santangelo, Computing counterion densities at intermediate
coupling, Phys. Rev. E \textbf{73}, 041512 (2006).

\bibitem {SL15}I. Y. Shilov and A. K. Lyashchenko, The role of concentration
dependent static permittivity of electrolyte solutions in the
Debye--H\"{u}ckel theory, J. Phys. Chem. B \textbf{119}, 10087-10095 (2015).

\bibitem {VB15}M. Valisk\'{o}, D. Boda, Unraveling the behavior of the
individual ionic activity coefficients on the basis of the balance of ion-ion
and ion-water interactions, J. Phys. Chem. B \textbf{119}, 1546 (2015).

\bibitem {VW16}J. H. Vera and G. Wilczek-Vera, \textit{Classical
Thermodynamics of Fluid Systems: Principles and Applications} (CRC Press, 2016).

\bibitem {V11}W. Voigt, Chemistry of salts in aqueous solutions: Applications,
experiments, and theory, Pure Appl. Chem. \textbf{83}, (2011) 1015-1030.

\bibitem {VB07}W. Voigt, et al., Quality assurance in thermodynamic databases
for performance assessment studies in waste disposal, Pure and applied
chemistry \textbf{79}, 883-894 (2007).

\bibitem {WR04}G. Wilczek-Vera, E. Rodil, and J. H. Vera, On the activity of
ions and the junction potential: Revised values for all data, AIChE. J.
\textbf{50}, 445 (2004).
\end{thebibliography}
\end{document}